\documentclass[varenna,seceqno]{cimento}
\usepackage{epsfig}
\title{ A statistical approach to the theory of the mean field
}
\author{ R. Caracciolo, A. De Pace \atque A. Molinari
}
\institute{ Dipartimento di Fisica Teorica dell'Universit\`a di
Torino and \\
Istituto Nazionale di Fisica Nucleare, Sezione di Torino, \\
 via P.Giuria 1, I-10125 Torino, Italy 
}
\author{ H. Feshbach
}
\institute{ Center for Theoretical Physics, \\
 Laboratory for Nuclear Science and Department of Physics, \\
 Massachusetts Institute of Technology, Cambridge, MA 02139, USA
}
\begin{document}

\maketitle

\section{Introduction}
\label{sec:intro}

The study of nuclear reactions has prompted the delineation of fundamental
principles whose importance extends beyond nuclear physics, from where they were
originally inferred, but reaches vast sectors of physics in general and perhaps
of other fields as well.

Nuclear reactions are of central interest because they deal with strong
interactions, for which perturbative approaches are of little value.
Indeed, these strong forces, acting at short distances between the nucleons,
entail large matrix elements to excited states of the nucleus lying at high
energy: this occurrence in turn leads to a poor convergence of the diagrammatic
expansion. In reaction theory the above problem finds a solution in the
recognition that once an average value (to be appropriately defined) is
subtracted from these matrix elements, then the resulting quantities are {\em
random}. Actually, we believe these results to hold not only for the continuum,
but for the nuclear bound states as well and also for other systems beside the
nuclear one. In other words we conjecture that many strongly interacting systems
in states of high excitations are {\em chaotic}.

In this conceptual framework an approach is here presented aiming at the
derivation of the nuclear mean field (MF), binding the nucleons together in the
nuclear ground state, from statistical considerations. In general, the mean
field happens to be a good description for a variety of many-body systems, whose
constituents can interact either with strong or weak forces, of long or short
range. It is thus natural to search for the features, common to the various
systems, ensuring the validity of the mean field approximation.
In line with the above speculations we hold the view that the statistical
aspect is the common feature linking all these different systems together.

Accordingly, we base our treatment of the MF on {\em two} propositions:
\begin{itemize}
\item[a)] the MF can be obtained through an energy average that smooths out 
short time events as, e.~g., those corresponding to the violent short-range 
nucleon-nucleon encounters in nuclei;
\item[b)] the matrix elements of the residual interactions, namely the force 
left out after the average interaction has been removed, are random with zero 
average value.
\end{itemize}

As we shall see these two elements suffice to obtain explicit expressions for
both the MF and the fluctuations (the {\em error}) away from the average.

Clearly the error should be small for the MF to be meaningful. In this
connection it turns out that, because of item b), the average error actually
vanishes. However, the average of the {\em square} of the error is not vanishing
and its magnitude we wish to assess. For this purpose, we expand the 
fluctuations
of the MF in terms of excitations of increasing complexities, each class of
excitations providing a contribution to the error. Notably, a general condition
for the convergence of this expansion (which is a finite one in the case of
atomic nuclei) can be derived; in fact, a parameter can be identified
determining the relative importance of each contribution to the error.

It is finally worth emphasizing that the theory here developed is, in the one
hand, quite general in the sense that no assumptions on the nature of the system
being studied are made, and, on the other, closely inspired from, in fact almost
identical with, the statistical theory of nuclear reactions.
Indeed it provides, as the latter does, an alternative method to account for the
contributions to physical observables stemming from processes slowly and rapidly
varying with the energy. It differs from the customary approaches which are
based on the explicit consideration of the Pauli and dynamical correlations, the
latter typically providing rapidly varying contributions.

In order to get an appreciation on how the method is practically implemented an
exploratory calculation of the ground state energy of an infinite, interacting
Fermi gas, i.~e. nuclear matter, will be discussed at the end of this lecture.
For this system a reasonable estimate of the error is obtained.
Furthermore, and significantly, the residual interaction appears to be
drastically reduced with respect to the bare one, an occurrence directly related
to the statistical nature of the present approach.

The present lecture, which is based on Ref.~\cite{Car97}, is organized as 
follows. In sect.~\ref{sec:formalism} the formalism is
revisited, --- in particular the partition of the Hilbert space into the $P$ and
$Q$ sectors, --- and the related equations are discussed, while in
{sect.}~\ref{sec:energy-average} a precise definition of the {\em energy
average} is provided. The latter is performed with a
distribution, characterized by a single parameter $\epsilon$, amounting to shift
the position of the poles corresponding to the excited states of the system away
from the ground state energy.
In the same section we pave the way for setting up a suitable formalism to
account for the corrections (the error) to the mean field energy.

In sect.~\ref{sec:corrections} the expression for the corrections to the mean
field energy $\bar{E}_0$ is given in the framework of the expansion, above
referred to, built out of successive 2p--2h excitations.

In sect.~\ref{sec:HF} we deal with the issue of introducing an operator,
projecting into a smooth $P$-space, simple enough to allow the present approach
to be worked out for nuclear matter. We thus restrict the $P$-space, --- also
with the aim of rendering transparent the comparison between the mean field
energy as given by the Hartree-Fock (HF) approach and by the present theory, 
--- just to a simple state: The HF
determinant. With this choice for $P$, we are able to establish an expression
connecting the true ground state $E$, the mean field $\bar{E}_0$ and the
Hartree-Fock
$E_{\sy{HF}}$ energies. Importantly, $\bar{E}_0$ is shown to be {\em always}
lower than $E_{\sy{HF}}$ and it turns out to be proportional to the variance,
to be later defined, of the residual effective interaction $V$ among nucleons,
which, in our context, is precisely defined.

In sect.~\ref{sec:spectr-fact} we address the issue of the
occupation probability $S^2$ (spectroscopic factor) of the ground state of
nuclear matter as defined by the structure of our $P$-space.
We are able to deduce, within our scheme, an expression for $S^2$ in terms of
$E$, $\bar{E}_0$ and $E_{\sy{HF}}$, and moreover embodying the energy
derivative of $V$.

Finally, in sect.~\ref{sec:numerical-results}
a scheme which permits numerical predictions of the
present approach is developed. It refers to the infinite Fermi gas, namely
nuclear matter.

For sake both of simplicity and of illustration, we
account for the first corrections only and, even so, a free parameter $\alpha$ 
should be introduced to obtain an estimate of their size.
Two systems of two equations each can then be set up by
coupling the equation for the mean field energy to the one for the fluctuations.
By solving the systems and with a convenient choice of the two parameters
entering into the present approach, namely $\epsilon$ and $\alpha$,
results consistent with the empirical features of nuclear matter, as derived
through an extrapolation from finite nuclei, are obtained.
At the same time an estimate for the fluctuations of the mean field energy is
provided.
These goals are achieved with a residual effective interaction which,
as previously emphasized, is lowered by about three orders of magnitude with
respect to the bare one
and yields a spectroscopic factor not in conflict with other independent
estimates \cite{Fant}.

In the concluding section the approach is summarized and possible improvements
discussed.

\section{Formalism}
\label{sec:formalism}

In this section we shortly present the formalism originally introduced in 
Ref.~\cite{Fes96}.
Let $H$ be the nuclear hamiltonian entering into the Schr\"odinger equation
\begin{equation}
 H \Psi = E \Psi  . 
\label{mascheq}
\end{equation}
Let $P$ be the hermitian operator projecting the nuclear ground state into the 
Hilbert subspace of functions associated with the low momentum transfer physics
and $Q$ the operator complementing $P$. Clearly
\begin{equation}
P^2 = P,\qquad Q^2 = Q,\qquad PQ = 0,\qquad P + Q = {\bf 1}  . 
\label{project}
\end{equation}
It is then proved that the pair of equations
\begin{equation}
(E - H_{PP} ) (P\Psi )= H_{PQ} (Q\Psi )  , 
\label{sist1}
\end{equation}
and 
\begin{equation}
(E - H_{QQ} ) (Q\Psi )= H_{QP} (P\Psi )  ,
\label{sist2}
\end{equation}
are equivalent to (\ref{mascheq}). In the above 
formulas the following shorthand
notations have been introduced, viz.
\begin{equation}
H_{PP} = PHP,\qquad H_{QQ} = QHQ,\qquad H_{PQ} = PHQ,\qquad
H_{QP} = QHP  .
\label{notaz}
\end{equation}
Now a direct way to proceed in order to obtain $P\Psi$ is to get from 
(\ref{sist1}) 
\begin{equation}
(P\Psi ) = \frac{1}{E - H_{PP}}  H_{PQ} (Q\Psi )  
\label{ppsieq}
\end{equation}
and from (\ref{sist2}) 
\begin{equation}
(Q\Psi ) = \frac{1}{E - H_{QQ}}  H_{QP} (P\Psi )  .
\label{qpsieq}
\end{equation}
Inserting then (\ref{qpsieq}) in (\ref{ppsieq}) one obtains
\begin{equation}
(P\Psi ) = \frac{1}{E - H_{PP}}  H_{PQ}\frac{1}{E - H_{QQ}} H_{QP}
 (Q\Psi )
\label{ppsi1eq}
\end{equation}
or, equivalently, after multiplication by $E-H_{PP}$, 
\begin{equation}
\left(E-H_{PP}-H_{PQ}\frac{1}{E-H_{QQ}}H_{QP}\right)(P\Psi) = 0 ,
\label{eq:eqsim}
\end{equation}
namely the equation obeyed by the component of $\Psi$ in the $P$-space.
A more general procedure is however required should a solution of the equation 
\begin{equation}
(E - H_{PP} ) \Psi_0 = 0 
\label{eqpsi0}
\end{equation}
exists.

In such a case, instead of eqs.~(\ref{ppsieq}) and (\ref{qpsieq}), one has 
\begin{equation}
(P\Psi ) = \Psi_0 + \frac{1}{E - H_{PP}}  H_{PQ} (Q\Psi )  
\end{equation}
and
\begin{equation}
(Q\Psi ) =\frac{1}{E - H_{QQ} - W_{QQ}} H_{QP} \Psi_0 \equiv
 \frac{1}{e_Q}  H_{QP} \Psi_0  ,
\end{equation}
where 
\begin{equation}
W_{QQ} = H_{QP} \frac{1}{E - H_{PP}} H_{PQ} 
\label{wqqdef}
\end{equation}
and
\begin{equation}
e_{Q} = E - H_{QQ} - W_{QQ}  .
\label{eqdef}
\end{equation}
The above equations allow then one to recast (\ref{sist1}) 
as follows
\begin{equation}
(E - H_{PP} ) (P\Psi )= H_{PQ}  \frac{1}{e_Q}  H_{QP} \Psi_0  
\label{sist1psi0}
\end{equation}
and to express $\Psi_0$ according to
\begin{equation}
\Psi_0 = \frac{1}{1 + 
\frac{\strut\displaystyle 1}{\strut\displaystyle E - H_{PP}} H_{PQ} 
\frac{\strut\displaystyle 1}{\strut\displaystyle e_Q} H_{QP}}
 (P\Psi )  .
\label{psi0eq}
\end{equation}
The combination of the two above equations leads in turn to
\begin{equation}
(E - H_{PP} ) (P\Psi )= H_{PQ}  \frac{1}{e_Q}  H_{QP} 
\frac{1}{1 + \frac{\strut\displaystyle 1}{\strut\displaystyle E - H_{PP}} 
  H_{PQ} \frac{\strut\displaystyle 1}{\strut\displaystyle  e_Q} 
H_{QP}}(P\Psi )  ,
\label{sist1exp}
\end{equation}
where no trace is left of $\Psi_0$.
Now using the operator identity
\begin{equation}
B \frac{1}{1 + CAB} = \frac{1}{1 + BCA} B ,
\label{opident}
\end{equation}
with the identifications 
\begin{equation}
A = \frac{1}{e_Q} , \qquad 
B = H_{QP}  ,\qquad C =  \frac{1}{E - H_{PP}}  H_{PQ}  ,
\label{abcdef}
\end{equation}
one obtains
\begin{eqnarray}
(E - H_{PP} ) (P\Psi ) & = & 
H_{PQ}  \frac{1}{e_Q} 
\frac{1}{1 + H_{QP} 
\frac{\strut\displaystyle 1}{\strut\displaystyle E - H_{PP}} H_{PQ} 
\frac{\strut\displaystyle 1}{\strut\displaystyle e_Q} }
H_{QP} (P\Psi )  \nonumber \\
 & = &  H_{PQ}  \frac{1}{e_Q} 
\frac{1}{1 + W_{QQ} 
\frac{\strut\displaystyle 1}{\strut\displaystyle e_Q}} H_{QP} (P\Psi )   .
\label{sist1exp1}
\end{eqnarray}
The obvious identity 
\begin{equation}
 \frac{1}{e_Q} \frac{1}{1 + W_{QQ} 
\frac{\strut\displaystyle 1}{\strut\displaystyle e_Q}}  = 
\frac{1}{\left (
\frac{\strut\displaystyle 1}{\strut\displaystyle e_Q}\right )^{-1}   
+ W_{QQ} }  
\label{1eqid}
\end{equation}
leads finally to 
\begin{equation}
\left (E - H_{PP} - H_{PQ} 
\frac{1}{\left (
\frac{\strut\displaystyle 1}{\strut\displaystyle e_Q}\right )^{-1}  
+ W_{QQ}} H_{QP}\right )
(P\Psi ) = 0  ,
\label{fesheqa}
\end{equation}
which is entirely equivalent to (\ref{eq:eqsim}).

It is important to note that the equation obeyed by $P\Psi$, being associated
with the intricate many-body operator 
\begin{equation}
{\cal H} = H_{PP}  + H_{PQ}
\frac{1}{\left (
\frac{\strut\displaystyle 1}{\strut\displaystyle e_Q}\right )^{-1}   
+ W_{QQ}} H_{QP} ,
\label{calh}
\end{equation}
is not an eigenvalue equation in the usual sense. 
Indeed, the dependence upon $E$, the exact ground state energy of the system, 
is non-linear, since the latter appears also in the propagator in the 
$Q$-space in (\ref{fesheqa}).

\section{ Averaging upon the energy }
\label{sec:energy-average}

The partition of the Hilbert space into a $P$ and a $Q$ sector should be
performed in conformity to the principle of including into the $P$-space the
wave functions with the simplest structure, namely those with a smooth spatial
dependence, and in the $Q$-space the wave functions of greater intricacy.
The actual $\Psi$ of the system should of course be viewed as a linear
superposition of components with all the possible degrees of complexity.

The problem then is: how to account for the average impact on the $P$-space 
of the $Q$-space wave functions, ignoring the detailed behaviour of the latter?
To deal with this question an averaging procedure should first be prescribed.
For this purpose we treat the energy $E$ as a variable upon which both the wave
function and the matrix elements depend. Clearly, the components of the wave
function varying most rapidly with $E$ lie in the $Q$-space: thus performing an
average over the energy amounts to smoothing out the behaviour of $Q\Psi$ which
can then be taken into account in the determination of the mean field.
The latter rules the gentle physics taking place in the $P$-space.

However, by replacing $(Q\Psi)$ with $\langle Q\Psi\rangle$ we also change at
the same time $(P\Psi)$ into $\langle P\Psi\rangle$ and $E$ into, say,
$\bar{E}_0$ (the angular brackets meaning, of course, energy averaging).

The system equivalent to the Schroedinger equation will now accordingly read 
\begin{equation}
\langle  P\Psi \rangle = \tilde \Psi_0 + \frac{1}{{\bar E_0} 
- H_{PP}}
  H_{PQ} \langle Q\Psi \rangle  ,
\label{avppsieq}
\end{equation}
where
\begin{equation}
\left ( {\bar E_0} - H_{PP} \right ) \tilde \Psi_0 = 0 
\label{phi0eq}
\end{equation}
and
\begin{equation}
\langle Q\Psi \rangle= \langle 
\frac{1}{e_Q}\rangle H_{QP} \tilde \Psi_0 .
\label{avqpsieq}
\end{equation}
Inserting (\ref{avqpsieq}) into (\ref{avppsieq}) yields 
\begin{equation}
\left ( {\bar E_0} - H_{PP} \right ) \langle P\Psi \rangle =
H_{PQ} \langle \frac{1}{e_Q}\rangle  H_{QP} \tilde \Psi_0  .
\label{avppsiphi0}
\end{equation}
Following then the same steps as before the equation
\begin{equation}
\left ( {\bar E_0} - H_{PP} 
- H_{PQ} \frac{1}{{\langle 
\frac{\strut\displaystyle 1}{\strut\displaystyle e_Q} \rangle}^{-1}  
+ W_{QQ}}  H_{QP}
\right )  \langle P\Psi \rangle  \equiv 
\left ( {\bar E_0} - {\bar {\cal H}}\right ) 
\langle P\Psi \rangle  = 0  
\label{avfesheq}
\end{equation}
is obtained: it clearly corresponds to equation (\ref{fesheqa}) when a suitable 
energy average has been performed.

We should now face the problem of specifying the energy average.
For this purpose a smoothing function $\rho (E , {\bar E_0} )$ with the property
\begin{equation}
\int \rho (E , {\bar E_0} )  dE = 1 
\label{intnew}
\end{equation}
should be introduced. 
Then averaging a function $f(E)$ means to perform the integral
\begin{equation}
\langle f \rangle  = \int \rho (E , {\bar E_0} )  f(E) dE 
\label{avef}
\end{equation}
which must be a real quantity since we are dealing with bound states.
A smoothing function obeying the above conditions is 
\begin{equation}
  \rho (E , \bar E_0 ) = \frac{1}{2\pi \sy{i}} 
     \frac{1}{E-(\bar E_0-\epsilon)} .
\label{rhoreal}
\end{equation}
In this case the integration in eq.~(\ref{avef}) is performed along a path 
coinciding with the real axis $Re E$ with a small
semi--circle described positively about the singularity 
$({\bar E_0} - \epsilon)$. 
If $f(E)$ is bounded at infinity sufficiently strongly, the Cauchy's integral 
formula can be applied so that the condition of eq.~(\ref{intnew}) is 
satisfied and eq.~(\ref{avef}) becomes 
\begin{equation}
\langle f \rangle  = f ({\bar E_0} -\epsilon) .
\label{avefeps}
\end{equation}
In the present approach $\epsilon$ is an empirical parameter essentially 
measuring the range over which the average is taken. 
In other words, for a sufficiently large value the energy of the 
states within $\epsilon$ can be neglected and it is in this sense that an 
average is performed. 

At this point the system corresponding to the original Schroedinger equation 
can be cast into the form (see also Kawai, Kerman and McVoy \cite{kaw75})
\begin{eqnarray}
(E - \bar {\cal H} ) (P\Psi ) & = &  V_{PQ} (Q\Psi )
\label{avsista} \\
(E - H_{QQ} ) (Q\Psi ) & = &  V_{QP} (P\Psi )
\label{avsistb}
\end{eqnarray}
where 
\begin{equation}
\bar {\cal H} = H_{PP} + V_{PQ} V_{QP}  
\frac{1}{\bar E_0  -\epsilon - E}  
\label{hbarv}
\end{equation}
and  
\begin{eqnarray}
V_{PQ} & = &  H_{PQ} \sqrt{\frac{\bar E_0 - \epsilon -E}
{\bar E_0  - \epsilon - H_{QQ}} }   , 
\label{vpqdefr} \\
V_{QP} & = &  \sqrt{\frac{\bar E_0  - \epsilon -E}
{\bar E_0  - \epsilon - H_{QQ}} }   H_{QP}  .
\label{vqpdefr}
\end{eqnarray}
Note that eqs.~(\ref{avsista}) and (\ref{avsistb}), while having the same
structure as the original eqs.~(\ref{sist1}) and (\ref{sist2}), display a
potential, coupling $(P\Psi)$ and $(Q\Psi)$, which is $V_{PQ}$ rather than 
$H_{PQ}$. 
The strength of the coupling is thus considerably reduced by roughly 
$(\epsilon / H_{QQ})^{\frac{1}{2}}$ which is much less than one. 
This in turn entails that $H_{PP}$ and $\bar {\cal H}$ are of the same order 
of magnitude.

Now  the spectral decomposition of the operator $1/( E - \bar {\cal H})$ can 
be performed in terms of eigenfunctions of the hermitian operator 
$\bar{\cal H}$, namely
\begin{equation}
\bar {\cal H} \Phi_n = \bar E_n \Phi_n .
\label{hbareig}
\end{equation}
It reads
\begin{eqnarray}
(P\Psi ) & = & \frac{1}{E - \bar {\cal H}} V_{PQ} (Q\Psi ) = 
\sum_{n} \frac{| \Phi_n \rangle \langle  \Phi_n |}{E - \bar E_n} 
V_{PQ} (Q\Psi ) \nonumber\\
& = & | \Phi_0 \rangle 
\frac{\langle \Phi_0 | V_{PQ}  | Q\Psi  \rangle}{E - \bar E_0} +
\left ( \frac {1}{E - \bar {\cal H}}\right )^\prime 
V_{PQ} (Q\Psi ) ,
\label{spectd}
\end{eqnarray} 
the prime on $(1/( E - \bar {\cal H}))$
signifying that the lowest eigenfunction $\Phi_0$ is to be excluded. 

From eq.~(\ref{spectd}) it follows
\begin{equation}
\langle \Phi_0 |  P\Psi   \rangle = 
\frac{\langle \Phi_0 | V_{PQ}  | Q\Psi  
\rangle}{E - \bar E_0}   .
\label{normc}
\end{equation}
To obtain $(Q\Psi )$ we return to eqs.~(\ref{avsista}) and (\ref{avsistb}) and 
get
\begin{equation}
(Q\Psi ) = \frac{1}{E - h_{QQ}} V_{QP} |\Phi_0\rangle 
\langle \Phi_0 | P\Psi \rangle  ,
\label{qpsiphi0}
\end{equation}
where 
\begin{equation}
h_{QQ} = H_{QQ} + \bar W_{QQ}  
\label{shqq}
\end{equation}
and 
\begin{equation}
\bar W_{QQ} \equiv V_{QP} \left ( \frac {1}{E - 
\bar {\cal H}}\right )^\prime
V_{PQ}  .
\label{wbardef}
\end{equation}
Inserting eq.~(\ref{qpsiphi0}) into eq.~(\ref{normc}) one finally obtains
\begin{equation}
 E - \bar E_0  = \langle \Phi_0 | V_{PQ}
\frac{1}{E - h_{QQ}} V_{QP} | \Phi_0 \rangle ,
\label{deltae}
\end{equation}
where the right hand side represents the correction to the mean field energy
$\bar{E}_0$. 

Remarkably, the quantity $\langle \Phi_0 | P\Psi \rangle$ does not appear in 
eq.~(\ref{deltae}), the reason being related to the arbitrariness of the
amplitude of $\Phi_0$. However, the bound state wave function $\Phi_0$ can be
normalized: our choice is $\langle \Phi_0 | P\Psi \rangle = 1$. 

\section{ Corrections to the energy of the mean field }
\label{sec:corrections}

We now search for an expression yielding the corrections to the ground state
energy $\bar{E}_0$ of the shell model in the statistical framework.
This is accomplished along the same line adopted in the theory of nuclear
reactions \cite{Fes92,Fes80}.
Accordingly (\ref{deltae}) is expressed through an expansion where each
contribution is not identified by a coupling constant to some power, but rather
by the degree of complexity of the states of the $Q$-space.

One thus writes
\begin{equation}
\Delta E = \sum_{m =1}^r \Delta E_m  ,
\label{deltaeexp}
\end{equation}
where
\begin{eqnarray}
\Delta E_m & = &  \langle \Phi_0 | V_{01} G_1 V_{12} G_2 
\cdots V_{m-1 , m} G_m V_{m,m-1} G_{m-1}
\cdots V_{10} | \Phi_0 \rangle  ,
\label{deltaem} \\
V_{ij} & = &  Q_i V Q_j  ,
\label{vij}
\end{eqnarray}
and $Q_j$ is the operator projecting into the $j$ particles--$j$ holes sector 
of the $Q$-space.

\begin{figure}
\mbox{\epsfig{file=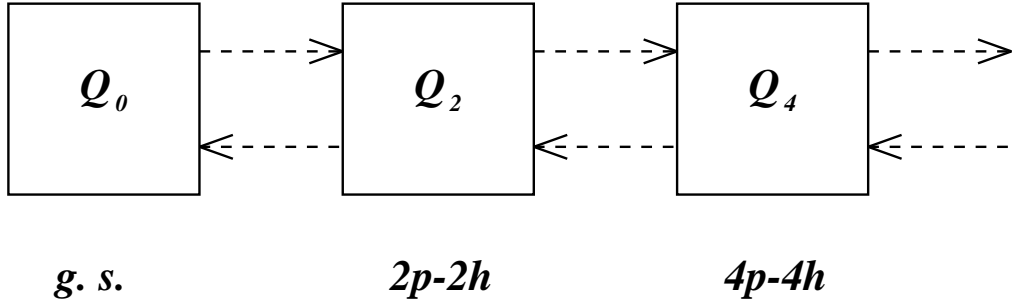}}
\vskip 3mm
\caption{
The partition of the Hilbert space of nuclear matter in sets of increasing
complexity. The first box on the left defines the $P$-space, the second one
embodies the simplest states in the $Q$-space and so on.}
\label{fig1}
\end{figure}

The above corresponds to a partition of the $Q$-space into $r$ sets, each one of
these embodying a specific class of excited states: thus to the $k$-th set are
associated the $k$ particle--$k$ hole excitations. Actually, assuming the
dominance of two-body forces in the residual interaction $V$ (the operator
connecting one set to the other), only excitations corresponding to an 
{\em even} number of particles
(and holes) are actually filling the sets, as indicated in Fig.~\ref{fig1}.
The propagation of the system into the $k$-th set is then described by the 
operator $G_k$, which, when $k=r$, reads
\begin{equation}
G_r = \frac{1}{E - h_{rr}}     ,
\label{grfin}
\end{equation}
whereas for $k < r$ obeys the recurrence relation
\begin{equation}
G_k = \frac{1}{E - h_{kk} - V_{k,k+1} G_{k+1} V_{k+1,k}} 
\label{gk}
\end{equation}
with
\begin{equation}
h_{kk} = H_{Q_k  Q_k} + V_{Q_k P} \left (\frac{1}{E - \bar {\cal H}}
\right )^\prime V_{P Q_k}  .
\label{hkkdef}
\end{equation}
Let us now, for convenience, pictorially describe a set as a box (see 
Fig.~\ref{fig1}): then eq.~(\ref{deltaem}) holds valid only if in the
expansion the occupation of a box occurs when the previous box has also been
occupied.

Notice also that in a nucleus the expansion
(\ref{deltaeexp}) is finite since the excited states with the greatest
complexity correspond to the case $r =A$, $A$ being the mass number.

Since the mean field contains the average, the average of the corrections to
the mean field vanishes. In addition because the processes occurring beyond 
the mean field are random, the average of the corrections to the
shell model arising from any individual set vanishes as well. 
Therefore we should have
\begin{equation}
\langle \Delta E \rangle = \langle \Delta E_1 \rangle + 
\langle \Delta E_2 \rangle + \cdots +
\langle \Delta E_A \rangle = 0 
\label{avdeltae}
\end{equation}
and 
\begin{equation}
\langle \Delta E_n \rangle = 0  .
\label{avdeltaen}
\end{equation}
However the average of the square of the corrections 
\begin{equation}
\langle (\Delta E)^2 \rangle =  \langle 
\sum_{n,m} \Delta E_n   \Delta E_m  \rangle 
\label{deltae2av}
\end{equation}
has no reason to vanish. Furthermore, owing to the 
randomness of the $Q$-space physics, the nuclear matrix elements entering into 
the definition of $\Delta E_n$ will also be random, hence the cancellation 
of the off-diagonal elements in (\ref{deltae2av}), leaving
\begin{equation}
\langle (\Delta E )^2 \rangle =  \langle
\sum_{n}(\Delta E_n )^2  \rangle  .
\label{deltae2avf}
\end{equation}

To ensure the fulfillment of the requirement
\begin{equation}
\langle \Delta E_1 \rangle = \langle \Delta E_2 \rangle = \cdots
= \langle \Delta E_A \rangle =0
\label{deltaei0}
\end{equation}
we redefine $\Delta E_1$ according to 
\begin{equation}
 \Delta E_1  = \langle \phi_A^0 | V_{01} G_1  V_{10}
 | \phi_A^0 \rangle - 
\left [ \langle \phi_A^0 | V_{01} G_1  V_{10}
 | \phi_A^0 \rangle \right ]_{AV}  ,
\label{avdeltae1}
\end{equation} 
and likewise for the terms with $r \not= 1$.
In (\ref{avdeltae1}) $\phi_A^0$ represents the nuclear mean field ground state 
wave function and the square brackets mean energy averaging.

We are now in a position to calculate, according to the formula (\ref{deltae}), 
the corrections to the mean field, provided the eigenfunctions 
$\psi_{k,\alpha}$ and the eigenvalues $\epsilon_{k,\alpha}$ of the operator 
$G^{-1}_k$ are known. We thus assume that the equation
\begin{equation}
  G^{-1}_k \psi_{k,\alpha} = (E-\epsilon_{k,\alpha}) \psi_{k,\alpha}
\end{equation}
can be solved for any value of the index $k$.

Let us then first consider $\Delta E_1$:
\begin{equation}
  \Delta E_1 
  = \sum_\beta\frac{|\langle\psi_{2\beta}|V|\phi_A^0\rangle|^2}
    {E-\epsilon_{2\beta}}
    - \left[\sum_\beta
    \frac{|\langle\psi_{2\beta}|V|\phi_A^0\rangle|^2}
    {E-\epsilon_{2\beta}}\right]_{AV}
     ,
\end{equation}
where $\epsilon_{2\beta}$ and $|\psi_{2\beta}\rangle$ are the eigenvalues and
eigenvectors, respectively, corresponding to the 2p--2h excitations.

Since in the present scheme only the square of the corrections to the mean 
field, namely
\begin{equation}
  (\delta E_1)^2 = \left[ (\Delta E_1)^2 \right]_{AV} - \left[ (\Delta E_1)
    \right]_{AV}^2  ,
\label{eq:deltaE1}
\end{equation}
are meaningful, we are left with the task of evaluating 
\begin{equation}
  (\delta E_1)^2 = \left[ \sum_{\beta\gamma}
  \frac{|\langle\psi_{2\beta}|V|\phi_A^0\rangle|^2 
       |\langle\psi_{2\gamma}|V|\phi_A^0\rangle|^2}
       {(E-\epsilon_{2\beta})(E-\epsilon_{2\gamma})} \right]_{AV} 
  - \left[ \sum_{\beta}\frac{|\langle\psi_{2\beta}|V|\phi_A^0\rangle|^2}
                           {E-\epsilon_{2\beta}} \right]^2_{AV}  .
\label{eq:deltaE1bis}
\end{equation}
Introducing an average excitation energy $\bar \epsilon_2$ for the 2p--2h 
states, the first term of the above equation can be recast as follows
\begin{eqnarray}
  &&  
  \left[ \sum_{\alpha, \beta}
    \frac{\langle \phi_A^0
    |V|\psi_{2\alpha}\rangle\langle\psi_{2\alpha}|V|\phi_A^0\rangle 
    \langle \phi_A^0|V|\psi_{2\beta}\rangle\langle\psi_{2\beta}|V|\phi_A^0
     \rangle}
         {(E-\epsilon_{2\alpha})(E-\epsilon_{2\beta})} \right]_{AV}
    \nonumber \\
  && \cong 
  \left( \frac{1}   {E-\bar \epsilon_{2}} \right)^2
\left[ \sum_{\alpha, \beta}
    \langle \phi_A^0
    |V|\psi_{2\alpha}\rangle\langle\psi_{2\alpha}|V|\phi_A^0\rangle 
    \langle \phi_A^0|V|\psi_{2\beta}\rangle\langle\psi_{2\beta}|V|\phi_A^0
     \rangle \right]_{AV} .
\end{eqnarray}
The random phase average of the quantity in the square brackets is next obtained
using 
\begin{equation}
 \langle A A^* B B^* \rangle = \langle A A^* \rangle
\langle B B^* \rangle  + \langle A^* B \rangle
 \langle A B^* \rangle + \sy{quartic terms} ;
 \label{aveab}  
\end{equation}
quartic terms will be neglected.

Now, the first term $\langle A A^* \rangle
\langle B B^* \rangle$ cancels exactly the second term in 
eq.~(\ref{eq:deltaE1bis}). The evaluation of the second term in 
eq.~(\ref{aveab}) then yields 
\begin{eqnarray}
  \left[ \left( \delta E_1 \right)^2 \right] &=& 
 \left( \frac{1}{E - \bar \epsilon_2}\right)^2 \sum_{\alpha}
    \left[ |\langle \phi_A^0
    |V|\psi_{2\alpha}\rangle|^2_{AV} \right]^2
    \nonumber \\
\label{avedeltae11}
  & \cong &
  \frac{\overline {\Delta E_1}}{(E - \bar \epsilon_2)^2}\frac{1}{D_2}
    \left[ |\langle \phi_A^0
    |V|\psi_{2}\rangle|^2_{AV} \right]^2 
\label{avedeltae1}
\end{eqnarray}
where $\overline {\Delta E_1}$, essentially given by $\epsilon$, is the 
interval over which the energy average is carried out. 
The quantity $D_2$ is the energy distance between two neighboring 2p--2h
states and, finally, $\psi_2$ is meant as a representative of the 2p--2h 
excitations.

We next consider $\Delta E_2$: 
\begin{eqnarray}
\Delta E_2 &=& 
 \langle \phi_A^0 | V_{01} G_1 V_{12} G_2 V_{21} G_{1}
V_{10} | \phi_A^0 \rangle  
\nonumber \\
&=& \sum_{\alpha, \beta, \alpha^\prime} 
\frac{\langle \phi_A^0
    |V|\psi_{2\alpha}\rangle}{E-\epsilon_{2\alpha}}
\frac{\langle \psi_{2\alpha}
    |V|\psi_{4\beta}\rangle}{E-\epsilon_{4\beta}}
\frac{\langle \psi_{4\beta}
    |V|\psi_{2\alpha^\prime}\rangle}{E-\epsilon_{2\alpha^\prime}}
 \langle \psi_{2\alpha^\prime}
    |V|\phi_A^0\rangle
    \nonumber \\
&\cong& \frac{1}{(E-\bar\epsilon_2)^2} \frac{1}{(E-\bar\epsilon_4)}  
\left[ \sum_{\alpha, \beta} \langle \phi_A^0
    |V|\psi_{2\alpha}\rangle
\langle \psi_{2\alpha}
    |V|\phi_A^0\rangle \right.
\nonumber \\
&& \cdot \langle \psi_{2\alpha}
    |V|\psi_{4\beta}\rangle \langle \psi_{4\beta}
    |V|\psi_{2\alpha}\rangle
 \nonumber \\
&& \left.   - \left[ \sum_{\alpha, \beta} |\langle \phi_A^0
    |V|\psi_{2\alpha}\rangle|^2
|\langle \psi_{2\alpha}
    |V|\psi_{4\beta}\rangle|^2 \right]_{AV} \right]
\end{eqnarray}
where we have set $\alpha = \alpha^\prime$, introduced an average energy 
$\bar{\epsilon}_4$ for the 4p--4h excitations and 
subtracted in the last line the average of $\Delta E_2$ as written in the 
first line in order to achieve $\langle \Delta E_2 \rangle = 0$. 
Focussing now on the square of the modified $\Delta E_2 = \delta E_2$ we get 
\begin{eqnarray}
(\delta E_2)^2  &\cong& 
\frac{1}{(E- \bar \epsilon_2)^4}\frac{1}{(E- \bar \epsilon_4)^2}\left[
 \sum_{\alpha, \beta, \alpha^\prime, \beta^\prime}
\langle \phi_A^0
    |V|\psi_{2\alpha}\rangle
\langle \psi_{2\alpha}
    |V|\phi_A^0\rangle \right.
\nonumber \\
&& \quad 
 \langle \psi_{2\alpha}
    |V|\psi_{4\beta}\rangle  \langle \psi_{4\beta}
    |V|\psi_{2\alpha}\rangle 
\bigg\{  \langle \phi_A^0
    |V|\psi_{2\alpha^\prime}\rangle  \langle \psi_{2\alpha^\prime}
    |V|\phi_A^0\rangle  
    \nonumber \\
&& \quad   \langle \psi_{2\alpha^\prime}
    |V|\psi_{4\beta^\prime}\rangle   \langle \psi_{4\beta^\prime}
    |V|\psi_{2\alpha^\prime}\rangle \bigg\} -  
 \sum_{\alpha, \beta, \alpha^\prime, \beta^\prime}
\left[ |\langle \phi_A^0
    |V|\psi_{2\alpha}\rangle |^2 \right.
\nonumber \\
&& \quad \left. \left.  |\langle \phi_A^0
    |V|\psi_{2\alpha^\prime}\rangle |^2
|\langle \psi_{2\alpha}
    |V|\psi_{4\beta}\rangle |^2 |\langle \psi_{2\alpha^\prime}
    |V|\psi_{4\beta^\prime}\rangle |^2 \right]_{AV} \bigg]_{AV} \right.  .
\end{eqnarray}
To calculate the average we again invoke eq.~(\ref{aveab}). As before
the first term cancels the subtracted term and one is left with 
\begin{equation}
(\delta E_2)^2 =
\frac{1}{(E- \bar \epsilon_2)^4}\frac{1}{(E- \bar \epsilon_4)^2}\left[
 \sum_{\alpha, \beta}
|\langle \phi_A^0
    |V|\psi_{2\alpha}\rangle|^4
|\langle \psi_{2\alpha}
    |V|\psi_{4\beta}\rangle|^4 \right]_{AV}  .
\label{eqdeltae2int}
\end{equation}
By comparing the above result with the one expressing 
$(\delta E_1)^2$, eq.~(\ref{avedeltae1}), it follows
\begin{equation}
(\delta E_2)^2 =
\frac{1}{(E- \bar \epsilon_2)^2}\frac{1}{(E- \bar \epsilon_4)^2}(\delta E_1)^2
|\langle \psi_{2}
    |V|\psi_{4}\rangle|^4
\label{eqdeltae2d1}
\end{equation}
from where an expression for the {\em expansion parameter} is emerging, the
expansion being however a finite one as previously emphasized.
Likewise $\psi_2$, in the above $\psi_4$ is describing a typical 4 particle--4
hole state. 

The generalization of eq.~(\ref{eqdeltae2d1}) to an expression connecting 
$(\delta E_n)^2$ and $(\delta E_{n-1})^2$ is then easily found to read 
\begin{equation}
\left[ (\delta E_n)^2\right]  =
\frac{1}{(E- \bar \epsilon_{2n})^2}\frac{1}{(E- \bar \epsilon_{2n-2})^2}
\left[ (\delta E_{n-1})^2 \right] 
\left[ |\langle \psi_{2n-2}
    |V|\psi_{2n}\rangle|^4\right]  .
\label{eqdeltae2dn}
\end{equation}

The results obtained up to this point have a general validity and can be applied
to any system thus justifying the concept of mean field: indeed they have been
deduced in the framework of the statistical theory. To further proceed a
detailed description of the system one wishes to study is required:
as an example we shall consider nuclear matter.

\section{ Explicit expressions for the projection operator and the HF theory }
\label{sec:HF}

We need now an explicit form for both the projection operator $P$
and the NN interaction in order to obtain the true energy ($E$) and the 
{\em mean field} one ($\bar E_0$) (the shell model energy for a nucleus).

First we search for a suitable projection operator. 
Suppose, for this purpose, that for a given interaction the HF problem has been
solved and let $\{ \phi_{\sy{HF}}^i \}$, with the 
index $i = 1,2, \dots$, be the single particle wave functions of the HF 
orbitals, which form an orthonormal complete set for the Hilbert space 
of a one particle system. Out of the $\{ \phi_{\sy{HF}}^i \}$ we can build an 
infinite set of Slater determinants $\{ \chi_{\sy{HF}}^i \}$  which also form 
an orthonormal complete set for the Hilbert space of the nucleus.

A natural choice for the projection operator is then 
\begin{equation}
P = \sum_{i=1}^M | \chi_{\sy{HF}}^i \rangle\langle  \chi_{\sy{HF}}^i | 
\label{pprojd}
\end{equation}
which, for $M=1$, reduces to 
\begin{equation}
P = | \chi_{\sy{HF}}^1 \rangle\langle  \chi_{\sy{HF}}^1 |   ,
\label{pproj1}
\end{equation}
$| \chi_{\sy{HF}}^1 \rangle$ being the ground state HF determinant of the 
nuclear system. 
In the following for sake both of simplicity and of illustration 
we shall stick to the case $M=1$, although generalizations 
to larger values of $M$, while cumbersome, should be worth exploring.

Notice that, with $P$ given by (\ref{pproj1}), the wave functions $(P\Psi )$ 
and $\langle P\Psi  \rangle$  turn out to be proportional to each other 
and to the ground state HF determinant $| \chi_{\sy{HF}}^1 \rangle$ as they 
must since the $P$ space has only one member.

With $P$ given by (\ref{pprojd}) and the mean field by $\bar {\cal H}$
(eq.~(\ref{hbarv})) one gets
\begin{equation}
  {\bar E_0} = E_{\sy{HF}} +
\langle \chi_{\sy{HF}}^1 |  VQQV^\dagger 
| \chi_{\sy{HF}}^1 \rangle 
\frac{1}{\bar E_0 -E -\epsilon}
\label{e0baravex}
\end{equation}
 where
\begin{equation}
E_{\sy{HF}} = 
\langle \chi_{\sy{HF}}^1 | H_{PP} | \chi_{\sy{HF}}^1 \rangle  . 
\label{ehfdef}
\end{equation}
and
\begin{equation}
  V = H \sqrt{\frac{\bar{E}_0-\epsilon-E}{\bar{E}_0-\epsilon-H_{QQ}}}
\label{eq:eff-int}
\end{equation}
(see eq.~(\ref{vpqdefr})).
Now writing writing $Q=1-P$ in Eq.~(\ref{e0baravex}), one obtains
\begin{equation}
\bar E_0 = E_{\sy{HF}} + 
\frac{1}{\bar E_0 - E -\epsilon}  
\bigg\{ \langle \chi_{\sy{HF}}^1 | V V^\dagger 
| \chi_{\sy{HF}}^1 \rangle -  \vert \langle \chi_{\sy{HF}}^1 | V
| \chi_{\sy{HF}}^1 \rangle\vert^2 \bigg\}  .
\label{e0barfin}
\end{equation}

Let us now specify the above general expression to the case of a Fermi gas 
(infinite nuclear matter), for which 
$|\chi_{\sy{HF}}^1 \rangle = | F \rangle$, $| F \rangle$ 
being the wave function of a Fermi sphere of radius $k_F$ (the Fermi 
wavenumber).
In infinite nuclear matter the quantity in curly brackets in 
eq.~(\ref{e0barfin}), which we refer to as the {\em variance} of the residual 
interaction $V$, simplifies considerably. Indeed is (the states 
$| n \rangle$ labelling the spectrum of the Fermi sphere)
\begin{eqnarray}
\langle F | V V^\dagger | F  \rangle & = &
\sum_n \langle F | V | n \rangle \langle n | V^\dagger | F  
\rangle \nonumber\\
 & \cong & \langle F | V | F  \rangle \langle F | V^\dagger | F \rangle
\nonumber\\
&  & + \sum_{1p-1h} \langle F | V | 1p-1h \rangle \langle
1p-1h  | V^\dagger | F \rangle \label{fv2fsv}\\
&  & +  \sum_{2p-2h} \langle F | V | 2p-2h \rangle \langle 
2p-2h  | V^\dagger | F \rangle  \nonumber\\
& = & \vert \langle F | V | F  \rangle \vert ^2 
+ \beta^2   , \nonumber
\end{eqnarray}
the piece related to the 1p-1h states essentially vanishing as required by the
Brillouin theorem \cite{Gro91} and further terms in the right hand side being
neglected because of the dominant two-body character of the residual
interaction $V$.
We indicate the variance of the residual interaction $V$, defined in 
eq.~(\ref{e0barfin}), with $\beta^2$ to remind that in nuclear matter such a
quantity turns out to be positive. To perform a reliable evaluation of $\beta^2$
is of course quite hard because the residual effective interaction is an
intricate operator: indeed, $V$ is not only energy-dependent, but dependent upon
the energy averaging procedure as well.
In addition it is highly-non-linearly connected to the hamiltonian acting in the
$Q$-space.

However, even in abeyance of an explicit expression for the variance, the
insertion (\ref{fv2fsv}) into (\ref{e0barfin}) leads to the important result 
\begin{equation}
\bar E_0 \cong E_{\sy{HF}} + 
\frac{\beta^2}{\bar E_0 - E -\epsilon}  ,
\label{e0barfin2}
\end{equation}
which embodies all the energies characterizing our problem, namely the HF, the 
mean field (which would correspond to the shell model in a finite nucleus) and 
the exact one.

From (\ref{e0barfin2}) one sees that even if a residual interaction $V$ is 
given and the associated HF energy
$E_{\sy{HF}}$ is calculated, still the energy $E$ of the system cannot be 
obtained since the mean field energy $\bar E_0$ remains to be fixed. 
However, if 
as a first orientation one sets $\bar E_0 \simeq E$, then the remarkable result
\begin{equation}
\bar E_0  \cong E_{\sy{HF}} - \frac{\beta^2}{\epsilon} 
\label{e0barappr}
\end{equation}
follows, showing that the mean field energy is {\em lower} than the HF 
one, the parameter $\epsilon$ being of course positive. 
This result holds only if $|\bar{E}_0-E| \ll \epsilon$.

\section{ Occupancy of the mean field ground state }
\label{sec:spectr-fact}

The projected wave function $P\Psi$ is just a component of the full wave
function $\Psi$: how big a fraction of $\Psi$ is contained in $P\Psi$ is
obviously a quantity of much interest. To estimate it let 
$\langle P\Psi|P\Psi\rangle = S^2$ and search for an equation for $S^2$ in
terms of the other quantities characterizing our problem. 
Since
\begin{equation}
  S^2 = 1 - \langle Q\Psi|Q\Psi\rangle 
\end{equation}
using eq.~(\ref{qpsiphi0}) for $(Q\Psi)$ one obtains
\begin{equation}
  S^2 = 1 - \langle \Phi_0|V_{PQ}\left(\frac{1}{E-h_{QQ}}\right)^2 V_{QP}
    |\Phi_0\rangle  .
\end{equation}
Take now the Fermi gas model for $\Psi$, i.~e. $|F\rangle$.
Then $\Phi_0=S|F\rangle$ and
\begin{equation}
  S^2 = 1 - S^2 \langle F|V_{PQ}\left(\frac{1}{E-h_{QQ}}\right)^2 V_{QP}
    |F\rangle 
\end{equation}
which, with some algebra and neglecting the weak energy dependence of $h_{QQ}$,
can be cast into the form
\begin{eqnarray}
  S^2 &\cong& 1 + S^2 \langle F|V_{PQ}\frac{d\phantom{E}}{dE}
    \left(\frac{1}{E-h_{QQ}}\right) V_{QP}|F\rangle \nonumber \\
  &=& 1 + S^2 \left[ \langle F|\frac{d\phantom{E}}{dE}\left(
    V_{PQ}\frac{1}{E-h_{QQ}}V_{QP}\right)|F\rangle \right. \nonumber \\
  &&\qquad - \left. \langle F|\frac{dV_{PQ}}{dE}\frac{1}{E-h_{QQ}}V_{QP}
    |F\rangle - \langle F|V_{PQ}\frac{1}{E-h_{QQ}}\frac{dV_{QP}}{dE}|F\rangle
    \right] .
\end{eqnarray}
Because
\begin{equation}
  \frac{dV_{PQ}}{dE} = -V_{PQ}\frac{1}{2(\bar{E}_0-\epsilon-E)}
\end{equation}
and
\begin{equation}
  \frac{dV_{QP}}{dE} = -V_{QP}\frac{1}{2(\bar{E}_0-\epsilon-E)}
\end{equation}
it follows that 
\begin{eqnarray}
  S^2 &=& 1 \nonumber \\
   && + S^2 \left[ \frac{d\phantom{E}}{dE}\langle F|V_{PQ}\frac{1}{E-h_{QQ}}
    V_{QP}|F\rangle + \frac{1}{\bar{E}_0-\epsilon-E}
    \langle F|V_{PQ}\frac{1}{E-h_{QQ}} V_{QP}|F\rangle \right]
\end{eqnarray}
or, by virtue of (\ref{deltae}), that
\begin{eqnarray}
  S^2 &=& 1 + S^2\left[\frac{d\phantom{E}}{dE}(E-\bar{E}_0) +
    \frac{E-\bar{E}_0}{\bar{E}_0-\epsilon-E}\right] \nonumber \\
  &=& 1 - S^2 \left( \frac{d\bar{E}_0}{dE} +
  \frac{\epsilon}{\bar{E}_0-\epsilon-E} \right)  .
\label{eq:S2Ebar0}
\end{eqnarray}
The explicit dependence of $\bar{E}_0$ upon $E$ is provided by (\ref{e0barfin2})
which, when inverted, yields
\begin{equation}
  \bar{E}_0 = \frac{1}{2}\left[E+\epsilon+E_{\sy{HF}}\pm
    \sqrt{(E+\epsilon-E_{\sy{HF}})^2 + 4\beta^2}\right]  ,
\end{equation}
the minus sign in front of the square root being taken, since for 
vanishing residual interaction ($\beta^2=0$) the mean field should reduce to 
the HF field ($\bar{E}_0=E_{\sy{HF}}$). Hence
\begin{equation}
  \frac{d\bar{E}_0}{dE} = \frac{1}{2} + \frac{1}{2} 
    \frac{E_{\sy{HF}}-E-\epsilon-2d\beta^2/dE}
         {\sqrt{(E_{\sy{HF}}-E-\epsilon)^2+4\beta^2}}  ,
\end{equation}
which, inserted into (\ref{eq:S2Ebar0}), finally leads to
\begin{eqnarray}
  S^2 &=& \left[ \frac{3}{2} + \frac{1}{2}
    \frac{E_{\sy{HF}}-E-\epsilon-2d\beta^2/dE}
         {\sqrt{(E_{\sy{HF}}-E-\epsilon)^2+4\beta^2}}
    +\frac{\epsilon}{\bar{E}_0-E-\epsilon}\right]^{-1} \nonumber \\
  &=& \left[ \frac{3}{2} + \frac{1}{2}
    \frac{E_{\sy{HF}}-E-\epsilon-2d\beta^2/dE}
         {\sqrt{(E_{\sy{HF}}-E-\epsilon)^2+4\beta^2}}
    + \epsilon\frac{\bar{E}_0-E_{\sy{HF}}}{\beta^2}\right]^{-1}  ,
\label{eq:S2constr}
\end{eqnarray}
namely to the equation we were looking for.
We thus see that the occupancy of the mean field ground state wave function 
$S^2$ is expressed in terms of the
HF, mean field and exact energies, with an additional dependence upon the energy
averaging parameter $\epsilon$, the residual interaction $\beta^2$ and the
derivative of the latter with respect to the energy.
Since all these quantities are either explicitly evaluated ($E_{\sy{HF}}$,
$\bar{E}_0$ and $E$) or fixed, in principle, both by the experiment ($\epsilon$)
and by our theoretical framework ($\beta^2$), eq.~(\ref{eq:S2constr}) provides
an important check for the consistency of our approach, also because
independent estimates on the depletion of the ground state wave 
function are available for lead, a nucleus for which nuclear matter is a 
reliable model, the present approach can be tested to some extent against the
experiment.

\section { A simple model for nuclear matter }
\label{sec:numerical-results}

To gain an insight into the effectiveness and self-consistency of the
formalism described above we develop here a simple, schematic model.
The model relies on two equations which, for sake of convenience, 
are here repeated. The first of these is 
eq.~(\ref{e0barfin2})

\begin{eqnarray}
&&\bar E_0 \cong E_{\sy{HF}} + 
\frac{\beta^2}{\bar E_0 - E -\epsilon}   .
\nonumber
\end{eqnarray}
The second is obtained from Eq.~(\ref{avedeltae11}), i.~e. 
\begin{eqnarray}
  \left[ \left( \delta E_1 \right)^2 \right] = 
 \left( \frac{1}{E - \bar \epsilon_2}\right)^2 \sum_{\gamma}
    \left[ |\langle \phi_A^0
    |V|\psi_{2\gamma}\rangle|^2_{AV} \right]^2   , 
 \nonumber
\end{eqnarray}
which we heuristically approximate (admittedly roughly) by 
\begin{equation}
 \left[ (\delta E_1)^2 \right]  = \frac{\beta^4}{(E-\alpha\epsilon)^2}   ,
 \label{eqst1}
\end{equation}
where $\alpha$ is a parameter which estimates the average energy of the 2p--2h
states involved (we set $\bar \epsilon_2 = \alpha \epsilon$ for convenience) 
and {\em importantly} helps to correct for the poor
approximation of $\beta^4$ for the sum over the 2p--2h states $\gamma$ in 
eq.~(\ref{avedeltae11}).
Taking the square root, from eq.~(\ref{eqst1}) it follows
\begin{equation}
  E-\bar{E}_0 = \pm \frac{\beta^2}{E-\alpha\epsilon}   .
\label{eq:fluct-guess}
\end{equation}

If now one combines eq.~(\ref{e0barfin2}) and (\ref{eq:fluct-guess}), using the 
upper
sign (plus) one obtains a lower bound $E_l$ for the energy of nuclear matter, 
while using the lower sign (minus) one obtains an upper bound $E_u$.
Clearly, one gets also differing values of the mean field energy to be denoted
by $\bar{E}_0^l$ and $\bar{E}_0^u$, respectively. We ask then whether there is a
choice of $\epsilon$, $\alpha$ and $\beta$ such that 
\begin{itemize}
\item[i)] the two mean field energies $\bar{E}_0^{l}$ and $\bar{E}_0^{u}$ 
turn out to be the same over a 
range of densities (or of Fermi momenta $k_F$) of significance for nuclear 
matter;
\item[ii)] the ``experimental'' values of the binding energy, saturation density
and compression modulus of nuclear matter, namely $B.E./A=-16$~MeV, 
$k_F=1.36$~fm$^{-1}$ and $K=23$ or 16 MeV (the two values refer to a hard and a 
soft equation of state, respectively) \cite{Bet71} are accounted for in a sense 
to be later 
specified;
\item[iii)] the spectroscopic factor given by eq.~(\ref{eq:S2constr}) turns out
to be less than one both on the 
lower bound (where its value $S_l$ is associated with $E_l$) 
and on the upper one (where its value $S_u$ is associated with $E_u$) and
furthermore not too much at variance with existing estimates.
\end{itemize}

To accomplish this program we proceed by first eliminating $\bar{E}_0^u$ and
$\bar{E}_0^l$ from eq.~(\ref{e0barfin2}) and (\ref{eq:fluct-guess}).
This yields an equation for $\beta_l^2$ and $\beta_u^2$:
\begin{equation}
  \beta^2_l = \frac{E_l-\alpha\epsilon}{2}\left\{[2E_l-\epsilon(\alpha+1)
    -E_{\sy{HF}}]-\sqrt{[2E_l-\epsilon(\alpha+1)-E_{\sy{HF}}]^2 
    +4\epsilon(E_l-E_{\sy{HF}})}\right\}   ,
\label{eq:beta2sola}
\end{equation}
for the lower bound, and
\begin{eqnarray}
  &&\beta^2_u = 
    \frac{E_u-\alpha\epsilon}{2}\left\{[E_{\sy{HF}}-\epsilon(\alpha-1)]
    -\sqrt{[E_{\sy{HF}}-\epsilon(\alpha-1)]^2 
    +4\epsilon(E_u-E_{\sy{HF}})}\right\}   ,
\label{eq:beta2solb}
\end{eqnarray}
for the upper bound.
Note that the solutions with the plus sign in front of the square root have been
discarded, since they lead to an incorrect limit as $\epsilon \to 0$.
The energy derivatives of (\ref{eq:beta2sola}) and (\ref{eq:beta2solb})are 
then easily obtained, namely
\begin{eqnarray}
  \frac{d\beta^2_l}{dE} &=& \frac{1}{2}\left\{[2E_l-\epsilon(\alpha+1)
    -E_{\sy{HF}}]-\sqrt{[2E_l-\epsilon(\alpha+1)-E_{\sy{HF}}]^2 
    +4\epsilon(E_l-E_{\sy{HF}})}\right\} \nonumber \\
&& \quad+(E_l-\alpha\epsilon)\left\{1-\frac{2E_l-\alpha\epsilon-E_{\sy{HF}}}
    {\sqrt{[2E_l-\epsilon(\alpha+1)-E_{\sy{HF}}]^2+
    4\epsilon(E_l-E_{\sy{HF}})}}
    \right\}   ,
\end{eqnarray}
for the lower bound, and
\begin{eqnarray}
  \frac{d\beta^2_u}{dE} &=& \frac{1}{2}\left\{[E_{\sy{HF}}-\epsilon(\alpha-1)]
    -\sqrt{[E_{\sy{HF}}-\epsilon(\alpha-1)]^2 
    +4\epsilon(E_u-E_{\sy{HF}})}\right\} \nonumber \\
  && \quad - (E-\alpha\epsilon)\frac{\epsilon}
    {\sqrt{[E_{\sy{HF}}-\epsilon(\alpha-1)]^2+4\epsilon(E_u-E_{\sy{HF}})}}
      ,
\end{eqnarray}
for the upper bound, These, when inserted into (\ref{eq:S2constr}), provide an
expression for the spectroscopic factor, associated with the lower and the 
upper bound, respectively, in terms of $E_l$ ($E_u$), $\bar{E}_0^l$ 
($\bar{E}_0^u$), $E_{\sy{HF}}$, $\alpha$ and $\epsilon$.

From eqs.~(\ref{eq:beta2sola}) and (\ref{eq:beta2solb}) we note that should 
appropriate values for
${E}_l$ (or ${E}_u$) and $E_{\sy{HF}}$ be available one would then obtain a 
value of $\beta_l^2$ and $\beta_u^2$ for given values of $\alpha$ and 
$\epsilon$.
These values of $\beta^2$ can in turn be used in eq.~(\ref{e0barfin2}) to obtain
$\bar{E}_0^l$ and $\bar{E}_0^u$. We agree that a solution of 
eq.~(\ref{e0barfin2}) and (\ref{eq:fluct-guess}) occurs when these upper and 
lower bounds for the mean field are equal.

The problem remains of how to choose meaningful values for $E_l$ and $E_u$ and 
how to fix $E_{\sy{HF}}$.
The first of these questions is answered by taking
\begin{eqnletter}
E_l &=& {\cal E}-W/2 \\
E_u &=& {\cal E}+W/2   ,
\end{eqnletter}
where the energy
\begin{equation}
  {\cal E} = [-16 +39.5(k_F-1.36)^2]  \sy{MeV} 
\label{eq:enuclmatt}
\end{equation}
incorporates the present knowledge on the ground state energy of nuclear matter,
quoted in ii), as extrapolated from finite nuclei and $W$ should be viewed as
the fluctuation energy.

\begin{figure}
\mbox{\epsfig{file=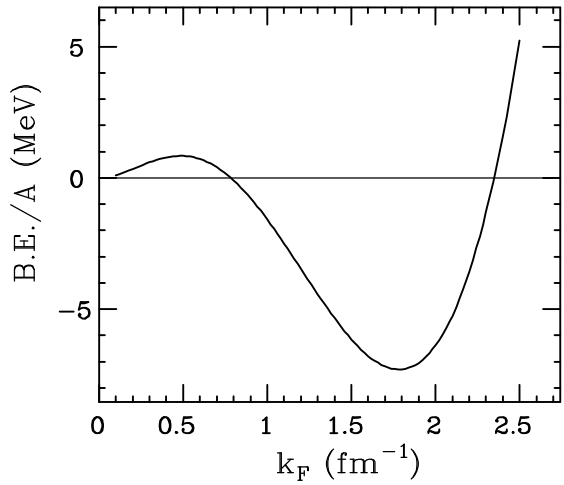}}
\vskip 3mm
\caption{
The binding energy per particle in the HF approximation (formula
(\protect\ref{binden}) of the text) for the potential (\protect\ref{nninter})
and the parameters given by (\protect\ref{gagbapp}).}
\label{fig2}
\end{figure}

The value of $E_{\sy{HF}}$ as a function of $k_F$ is obtained from an assumed,
schematic 
two-body interaction (Appendix~\ref{app:app}).
The resulting Hartree-Fock energy for a Fermi gas is shown in Fig.~\ref{fig2}
and, for a reduced scale in $k_F$, in Fig.~\ref{fig4}.
From Fig.~\ref{fig2} we see that for our simple two-body force a
minimum of the binding energy as a function of $k_F$ occurs at $k_F=1.78$ 
fm$^{-1}$, the corresponding energy per particle being $-$7.3 MeV.

The potential can also be used to calculate the bare value of $\beta^2$ 
(Appendix~\ref{app:appb}). Obtained values are shown in Tables
\ref{tab:beta2_2}, \ref{tab:beta2_3} and \ref{tab:beta2_4}:
they range from $4.5\times10^4$ to $7.0\times10^4$ MeV$^2$/nucleon.

\begin{figure}
\mbox{\epsfig{file=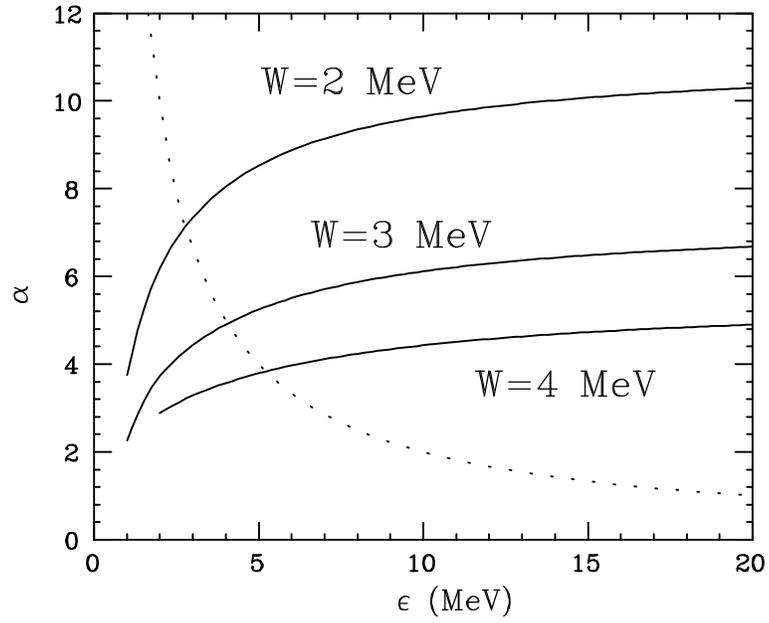}}
\vskip 3mm
\caption{ The loci corresponding to $\bar{E}_0^l=\bar{E}_0^u$ in the plane
($\alpha,\epsilon$), for $W=2$, 3 and 4 MeV. Also shown (dotted line)
is the curve along which the average energy of the 2p--2h excitations is  20 MeV
}
\label{fig3}
\end{figure}

\begin{figure}
\mbox{\epsfig{file=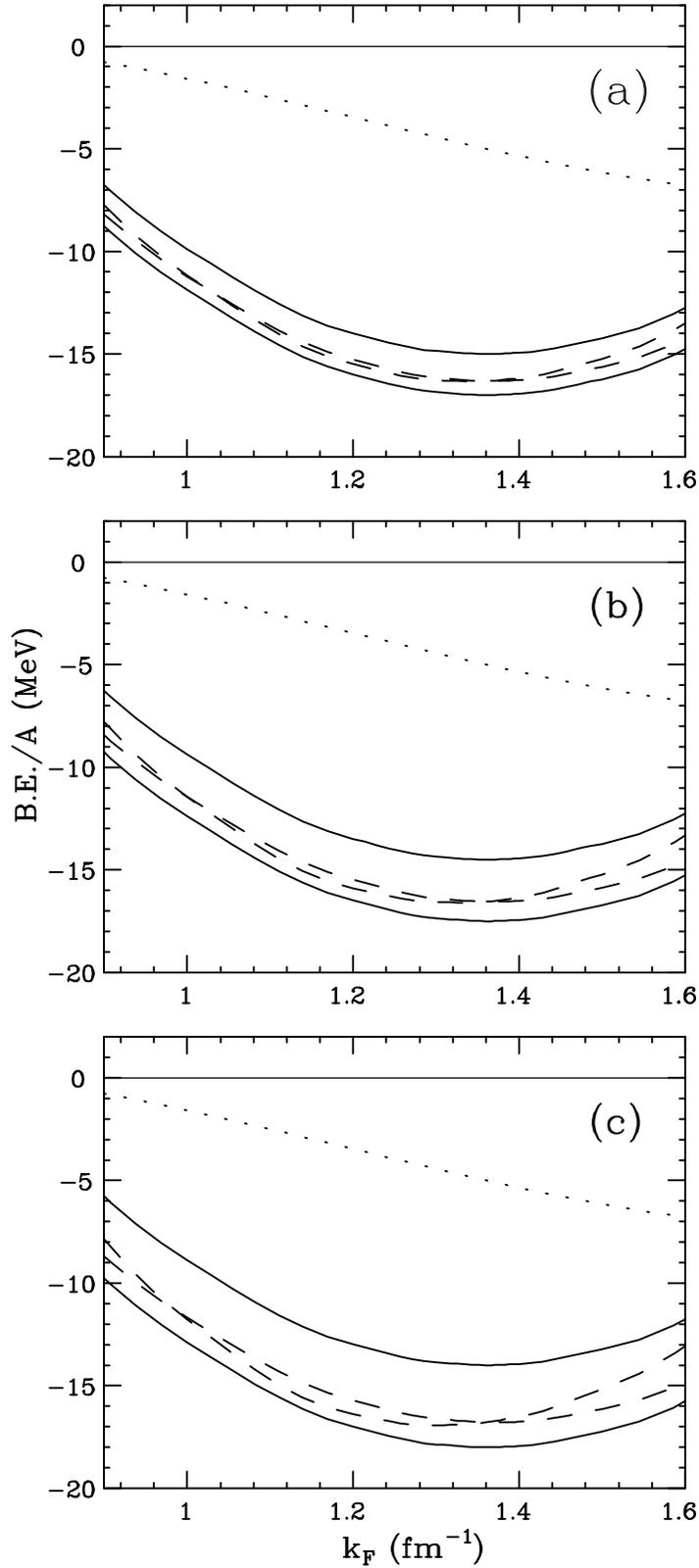}}
\vskip 3mm
\caption{ The binding energy per particle in the HF approximation (dot) and in
the present approach: The solid lines represent the lower and upper bounds of
fluctuations for the energy $E$, whereas the dashed lines give the corresponding
mean field energies.
(a): $W=2$ MeV, $\epsilon=3$ MeV, $\alpha=7.3$, $\bar{\epsilon}_2 = 22$ MeV;
(b): $W=3$ MeV, $\epsilon=4$ MeV, $\alpha=4.9$, $\bar{\epsilon}_2 = 20$ MeV;
(c): $W=4$ MeV, $\epsilon=5$ MeV, $\alpha=3.8$, $\bar{\epsilon}_2 = 19$ MeV.
}
\label{fig4}
\end{figure}

In discussing our findings it help first to observe that actually, for a given 
fluctuation energy
$W$, a whole set of values for the parameters $\alpha$ and $\epsilon$ 
exist such to satisfy the requirement i), namely 
$\bar{E}_0^l =\bar{E}_0^u$, at $k_F =1.36$ fm$^{-1}$.
They lie on the curves displayed in Fig.~\ref{fig3}. 
Furthermore $\alpha$ and $\epsilon$ should also be such to fulfill the 
constraint
\begin{equation}
\bar \epsilon_2 = \alpha \epsilon \ge 20 {\sy { MeV}}   ,
\end{equation}
which represents a fair estimate of the lower limit 
for the excitation energy of the 2p--2h states.
As a consequence the acceptable values for $\alpha$ and $\epsilon$ are then 
restricted to the domain to the right of the dotted line in 
Fig.~\ref{fig3}.

In this region $\alpha$ and $\epsilon$  should be selected in such a way to
comply with the requirements ii) and iii).
This turns out to be possible and our results are shown in 
 Fig.~\ref{fig4} and Tables \ref{tab:beta2_2}, \ref{tab:beta2_3} and
\ref{tab:beta2_4}.

However there one sees that as one moves away from $k_F =1.36$ fm$^{-1}$
the equality between $\bar{E}_0^l$ and  $\bar{E}_0^u$
is no longer exactly satisfied, although the two mean fields remain rather close
to each other at least in the range $0.9 \le k_F \le 1.5$  fm$^{-1}$
providing the values of $W$ are ``moderate''.
Indeed, quantitatively, requirement i) can be reasonably satisfied in the above
quoted range of $k_F$ for $W \le 4$ MeV only. For larger fluctuation energies 
$\bar{E}_0^l$ and  $\bar{E}_0^u$ differ too much: an orientation on the size of
the error around the mean field in thus obtained.

Concerning the energy averaging parameter, $\epsilon$, it turns out to be about
1 MeV larger that $W$ for all
the cases listed in the tables. The values of the spectroscopic factors 
$S_l$ and $S_u$ are quite stable, while the values of the effective interaction
$\beta_l^2$ and $\beta_u^2$ grow rapidly with increasing $W$. Most importantly, 
their values differ by {\em three orders of magnitude} from the bare
$\beta^2$. This occurrence partly stems from the renormalization (see 
eqs.~(\ref{vpqdefr}) and (\ref{vqpdefr})) induced by the energy averaging.
The random phase averaging also gives rise to a reduction in the magnitude of
the residual interaction. Calculations, while rough, still indicate that these 
two effects appear indeed sufficient to produce the observed sharp reduction.

\section{Closing comments}
\label{sec:conclusions}

The statistical theory of the mean field presented in sections 
\ref{sec:intro}--\ref{sec:corrections} is based on two propositions. 
It is assumed that the mean field is the slowly
varying component of the nuclear interaction, which can be obtained by taking an
appropriate energy average. Secondly it is suggested that the matrix elements of
the residual interaction are random, so that their average is zero. A formalism
incorporating these ideas, borrowed from statistical reaction theory, was 
developed and explicit expressions for the mean field and the ``fluctuation''
away from the mean field was obtained. 

Next an expansion of the fluctuation energy in terms of
increasing excitation complexity leads, after averaging, to formulas for the
corresponding contributions to fluctuation energy.
Worth emphasizing is the novelty of this approach: indeed the corrections to 
the mean field ground state energy of the system rather than being ordered 
according to the
power of a coupling constant or according to the number of hole lines occurring
in a given diagram, as it is done in standard theories, are here organized in 
terms of the complexity of the $Q$-space states reached via the residual 
interaction acting among the constituents of the system.

The presentation of this approach was followed in sect.~\ref{sec:HF},
\ref{sec:spectr-fact} and \ref{sec:numerical-results} by a simplified version 
and by a schematic model. The most remarkable result emerging from the analysis 
of the model has been the sharp reduction of the
effective strength of the residual interaction. But in addition is the overall
reasonableness of our results which is quite encouraging.

Of course much remains to be done. The quantitative connection with the 
underlying
nuclear forces has not been exhibited. The evaluation of the matrix elements
for finite  nuclei was not carried out and a better understanding of the 
energy average needs to be achieved. We have so far only a schematic model. What
is needed is a complete and thorough evaluation which, on the basis of the 
obtained results can be expected with confidence to be successful.

Application to excited states is also indicated. In this case the smoothing
function used in reaction theory can be used instead of the one of section
\ref{sec:corrections}. This would lead to a complex mean field and the excited
states would have a width corresponding to the probability of the splitting of 
the state by the residual interaction. The width would then measure the extent 
of the splitting.

Of crucial importance is also to test the rate of convergence of the 
complexity expansion. Three elements are expected to favour in general a fast 
convergence rate for the latter: 
\begin{itemize}
\item[i)] the energy averaging parameter $\epsilon$, if it is large enough;
\item[ii)] the subtraction of the average interaction from the bare one;
\item[iii)] the reduced overlap between the ground state and the complex wave 
functions belonging to the $Q$-space.
\end{itemize}

A detailed investigation will be invaluable in sheding light on this 
issue. In this connection calculations are presently in progress.

\appendix
\section {The bare interaction and the HF theory}
\label{app:app}

To implement the program outlined in Section \ref{sec:numerical-results}
we need a NN interaction to fix the HF energy of
nuclear matter.
 For illustrative purposes we choose the following simple NN interaction
\begin{equation}
 {\cal V}(r) = g_A   \frac{e^{-\mu_A r}}{r}  - 
   g_B  \frac{e^{-\mu_B r}}{r}   \frac{ 1 + P_x }{2}    ,
\label{nninter}
\end{equation}
which embodies a short-range repulsion in the first term and an intermediate 
range attraction in the second one (all the parameters are positive and
we require $\mu_A > \mu_B$).
The latter is taken of Majorana type, hence the occurrence of
the exchange operator
\begin{equation}
P_x = \frac{1 + \vec{\tau}_1\cdot\vec{\tau}_2}{2} 
\frac{1 + \vec{\sigma}_1\cdot\vec{\sigma}_2}{2} 
\label{pxmajo}
\end{equation}
built out of the spin $\vec{\sigma}$
and isospin $\vec{\tau}$ operators. Thus (\ref{nninter})
contains the main features needed to account for the saturation
of the nuclear forces, reflected in  the existence of a minimum 
in the binding energy per particle $(B.E./A)$ versus $k_F$ curve,
but for the tensor force that here, for the sake of simplicity, is neglected.

The HF energy for the interaction  (\ref{nninter}) is easily worked out and 
leads to the following expression for the binding energy per particle 
\cite{Alb80}
\begin{eqnarray}
  \frac{B.E.}{A} & = & \frac{3}{5} \frac{\hbar^2 k_F^2}{2m} + 
    \frac{\varrho}{2}  \left ( 4\pi\frac{g_A}{\mu_A^2} - 
    \frac{3\pi}{2} \frac{g_B}{\mu_B^2} \right ) 
     - \frac{3k_F}{4\pi} \left ( g_A + \frac{3}{2} g_B \right ) 
    \nonumber\\
 &&+\frac{1}{8\pi k_F} \left ( g_A \mu_A^2 + \frac{3}{2} g_B \mu_B^2 \right )
    \nonumber\\
  & & +\frac{1}{\pi} \left [  g_A \mu_A   
    \arctan \left (\frac{2k_F}{\mu_A}\right )
    + \frac{3}{2}   g_B \mu_B    \arctan \left (\frac{2k_F}{\mu_B}\right ) 
    \right ] \label{binden} \\
  & & - \frac{1}{8\pi k_F} \Biggl [  g_A \mu_A^2 
    \left( 3 + \frac{\mu_A^2}{4k_F^2}\right )
      \log \left ( 1 + \frac{4 k_F^2}{\mu_A^2}\right ) 
    \nonumber\\
  && 
    + \frac{3}{2}   g_B \mu_B^2  \left( 3 + \frac{\mu_B^2}{4k_F^2}\right )
      \log \left ( 1 + \frac{4 k_F^2}{\mu_B^2}\right )\Biggr ]   ,
\nonumber
\end{eqnarray}
where $\varrho = {2 k_F^3}/{3\pi^2}$.

The parameters characterizing the potential (\ref{nninter}) might be fixed,
for example, by accounting for the ``experimental'' nuclear matter values 
previously quoted.
One succeeds in doing so with the following choice
\begin{eqnletter}
\mu_A & = &  3.43  \sy{fm}^{-1}   ,\qquad \qquad  
\mu_B =   1.63  \sy{fm}^{-1}
\label{muamub} \\
g_A&  = & 2460  \sy{MeV~fm}  ,\qquad    g_B  =  898 
 \sy{MeV~fm}  .
\label{gagbexa}
\end{eqnletter}

Worth noticing is that the range of the repulsion obtained with the 
fitting procedure is rather close to the one associated with the exchange of a 
$\omega$ meson  (3.97 fm$^{-1}$), whereas the range of the attraction
turns out to be intermediate to the one arising from the exchange of a pion
and of a $\sigma$ meson (0.71 fm$^{-1}$ and 2.79 fm$^{-1}$, respectively).

Here, we rather prefer to choose the parameters in such a way to have too 
little binding energy at too large a density in the HF frame, in order to 
conform to a shortcoming common to many nuclear matter calculations, the 
purpose being to ascertain whether the present theory is capable to improve 
upon the HF results.
Of course, there exists a variety of ways for reaching this scope:
In view of the rather realistic values of the ranges $\mu_A$ and $\mu_B$
(see (\ref{muamub})) we change the coupling constants.
We thus take, as a rather extreme  example,
\begin{equation}
g_A = 740 \sy{MeV~fm}   ,\qquad g_B = 337  \sy{MeV~fm}  ,
\label{gagbapp}
\end{equation}
which, together with the values for $\mu_A$ and $\mu_B$ given in (\ref{muamub}),
yields a minimum of $-7.30$ MeV for the binding energy at $k_F=1.78$ fm$^{-1}$, 
as it can be seen in Fig.~\ref{fig2}, where the HF energy is displayed as a 
function of the density.

\section{Vacuum $\to$ 2p--2h matrix element}
\label{app:appb}

In the language of second quantization the matrix element we have to calculate
reads
\begin{equation}
 \beta^2 = \sum_{\scriptstyle\sy{spin}\atop\scriptstyle\sy{isospin}}
   \sum_{\scriptstyle k_1,k_2<k_F\atop
         \scriptstyle|\vec{k}_1+\vec{q}|,|\vec{k}_2-\vec{q}|>k_F}
  \langle\vec{k}_1,\vec{k}_2|V|\vec{k}_1+\vec{q},\vec{k}_2-\vec{q}\rangle
  \langle\vec{k}_1+\vec{q},\vec{k}_2-\vec{q}|V|\vec{k}_1,\vec{k}_2\rangle,
\label{eqb1}
\end{equation}
which, with standard manipulations, can be transformed into
\begin{eqnarray}
\beta^2 & = & A \frac{2}{\pi \varrho}
\int \frac{d \vec{k}_1}{(2\pi)^3}\frac{d \vec{k}_2}{(2\pi)^3} d \vec{q}
      \Theta \left ( |\vec{k}_1 +\vec{q} | -k_F \right )
\Theta \left ( |\vec{k}_2 -\vec{q} | -k_F \right )
\Theta \left ( k_F -k_1 \right ) \nonumber \\
&  &  \times\Theta \left ( k_F -k_2 \right ) 
\Biggl \{ 32 g_A^2 \frac{1}{(\mu_A^2 +q^2 )^2} +
12 g_B^2 \frac{1}{(\mu_B^2 +q^2 )^2}  
- 24 g_A g_B \frac{1}{\mu_A^2 +q^2}   
\frac{1}{\mu_B^2 +q^2} \nonumber \\
&  & - 8 g_A^2 \frac{1}{\mu_A^2 +q^2}   \frac{1}{\mu_A^2 +| \vec{k}_1 - 
\vec{k}_2 + \vec{q} |^2} 
+  12 g_B^2 \frac{1}{\mu_B^2 +q^2}   \frac{1}{\mu_B^2 +| \vec{k}_1 - 
\vec{k}_2 + \vec{q} |^2} \label{beta2} \\
&  &  -  12 g_A g_B \left ( \frac{1}{\mu_A^2 +q^2}    
\frac{1}{\mu_B^2 +| \vec{k}_1 - \vec{k}_2 + \vec{q} |^2} + 
\frac{1}{\mu_B^2 +q^2}   \frac{1}{\mu_A^2 +| \vec{k}_1 -
\vec{k}_2 + \vec{q} |^2} \right ) \Biggr \}   \nonumber 
\end{eqnarray}
(the matrix elements in eq.~(\ref{eqb1}) are understood to be antisymmetrized).
The above 9-dimensional integral,
by appropriate transformations, can be
reduced to a combination of 2- and 4-dimensional integrals, which can be
numerically evaluated, yielding the results quoted in  
Tables~\ref{tab:beta2_2}--\ref{tab:beta2_4}.
These values turn out to be helpful in performing a comparison with later 
findings.

The direct contributions, namely the first three terms in the right hand side of
(\ref{beta2}), are analytically evaluated.
Considering, e.~g., the third one, one gets 
\begin{eqnarray}
A \frac{2}{\pi \varrho} &&
\int \frac{d \vec{k}_1}{(2\pi)^3}\frac{d \vec{k}_2}{(2\pi)^3} d \vec{q}
      \Theta \left ( |\vec{k}_1 +\vec{q} | -k_F \right )
\Theta \left ( |\vec{k}_2 -\vec{q} | -k_F \right )
\Theta \left ( k_F -k_1 \right ) \Theta \left ( k_F -k_2 \right )
\nonumber \\
&&\qquad\times\left\{ \frac{1}{\mu_A^2 +q^2}   
\frac{1}{\mu_B^2 +q^2} \right\} \nonumber \\
&& = A \frac{\varrho}{4 k_F} \Biggl\{ \frac{1}{\tilde \mu_A^2 - \tilde \mu_B^2} 
\left (\tilde \mu_A \arctan \tilde \mu_A -\tilde \mu_B \arctan \tilde \mu_B
\right ) \\
&& + \frac{9}{5} + \frac{17}{12} (\tilde \mu_A^2 + \tilde \mu_B^2 )
+ \frac{1}{4} (\tilde \mu_A^4 + \tilde \mu_B^4 ) +
\frac{1}{4} \tilde \mu_A^2  \tilde \mu_B^2 \nonumber \\
&& +\frac{1}{4} \frac{1}{\tilde \mu_A^2 - \tilde \mu_B^2}  \left [
 \tilde \mu_B^3 (3 + \tilde \mu_B^2 )^2 \arctan 
\left ( \frac{1}{\tilde \mu_B} \right ) 
- \tilde \mu_A^3 (3 + \tilde \mu_A^2 )^2 \arctan
\left ( \frac{1}{\tilde \mu_A} \right )
\right ] \Biggr\} \nonumber   ,
\end{eqnarray}
where $\tilde \mu_{A,B} = {\mu_{A,B}}/{2 k_F}$.

The other terms are obtained with obvious substitutions.

The exchange contribution is somewhat more involved. As an example we consider 
(again the other pieces are obtained through obvious interchanges)
\begin{eqnarray}
  \beta^2_{\sy{exc}} &=& A \frac{2}{\pi \varrho} 
\int \frac{d \vec{k}_1}{(2\pi)^3}\frac{d \vec{k}_2}{(2\pi)^3} d \vec{q}
    \nonumber\\
    &&  \Theta \left ( |\vec{k}_1 +\vec{q} | -k_F \right )
\Theta \left ( |\vec{k}_2 -\vec{q} | -k_F \right )
\Theta \left ( k_F -k_1 \right ) \Theta \left ( k_F -k_2 \right )
\nonumber \\
&&\qquad\times\left\{ \frac{1}{\mu_A^2 +q^2}  
\frac{1}{\mu_B^2 + | \vec{k}_1 - \vec{k}_2 + \vec{q} |^2}
 \right\} \nonumber \\
&=& \frac{A}{(2 \pi^2 )} \frac{2}{\pi^2 \varrho} \sum_{l=0}^\infty 
(- 1)^l (2l +1) \int_0^\infty dq  \frac{q^2 }{\mu_A^2 +q^2}
\int_0^\infty dr  r e^{-\mu_B r} j_0 (qr) \nonumber   \\
&& \qquad \times \left [ \int_0^{k_F} dk  k^2 j_l (kr) 
\int_{-1}^1 dx  P_l (x)   \Theta (q^2 + k^2 + 2 q k x - k_F ) 
\right ]^2 , 
\end{eqnarray}
where a partial wave expansion, in terms of the Legendre polynomials $P_l$ and
the spherical Bessel functions $j_l$, has been performed. 
The above integral, for $q\ge 2k_F$ is easily calculated yielding
\begin{eqnarray}
  \beta^2_{\sy{exc}} &=& A \frac{2}{\pi^4 \varrho}
\int_0^\infty dr  \frac{e^{-\mu_B r}}{r^6}  
\Bigl ( \sin (k_F r) - k_F r \cos (k_F r) \Bigr )^2 \nonumber \\
 && \times \int_{2 k_F}^\infty dq  \frac{q }{\mu_A^2 +q^2}
\sin (qr)   .
\end{eqnarray}
The remaining integrals are numerically evaluated.

\begin{table}
\caption{ In the second row the value of $\bar{E}_0$ as a function of $k_F$ is
reported; in the third row one finds the bare value of $\beta^2$, as given by
eq.~(\protect\ref{beta2}), while in the fourth and fifth rows the
estimated renormalized values on the lower and upper bounds, respectively;
in the last two rows, the corresponding values of the spectroscopic factor are
displayed. The values in this table correspond to $W=2$ MeV, $\epsilon=3$ MeV
and $\bar{\epsilon}_2=22$ MeV.
 }
\label{tab:beta2_2}
\begin{tabular}{lccc}
  \hline
      $k_F$ (fm$^{-1}$) & 1.2  & 1.36 & 1.5 \\
  \hline
  $\bar{E}_0$ (MeV)   & -15.3$\div$-15.5 & -16.3  & -15.6$\div$-15.2 \\
  \hline
  $\beta^2$   & 4.5$\cdot$10$^4$  & 5.2$\cdot$10$^4$  & 7.0$\cdot$10$^4$  \\
  $\beta^2_l$ (MeV$^2$/nucleon) & 27.0  & 26.3  & 22.8  \\
  $\beta^2_u$                   & 54.2  & 49.0  & 36.4  \\
  \hline
  $S_l$       &  0.32  &  0.32  &  0.30  \\
  $S_u$       &  0.64  &  0.62  &  0.58  \\
  \hline
\end{tabular}
\vskip 3mm
\caption{ As in table~\protect\ref{tab:beta2_2}:
The values in this table correspond to $W=3$ MeV, $\epsilon=4$ MeV
and $\bar{\epsilon}_2=20$ MeV.
 }
\label{tab:beta2_3}
\begin{tabular}{lccc}
  \hline
      $k_F$ (fm$^{-1}$) & 1.2  & 1.36 & 1.5 \\
  \hline
  $\bar{E}_0$ (MeV)   & -15.5$\div$-15.9 & -16.6  & -15.9$\div$-15.2 \\
  \hline
  $\beta^2$   & 4.5$\cdot$10$^4$  & 5.2$\cdot$10$^4$  & 7.0$\cdot$10$^4$  \\
  $\beta^2_l$ (MeV$^2$/nucleon) & 36.1  & 35.3  & 30.7  \\
  $\beta^2_u$                   & 79.6  & 69.9  & 49.8  \\
  \hline
  $S_l$       &  0.34  &  0.33  &  0.32  \\
  $S_u$       &  0.69  &  0.67  &  0.63  \\
  \hline
\end{tabular}
\vskip 3mm
\caption{ As in table~\protect\ref{tab:beta2_2}:
The values in this table correspond to $W=4$ MeV, $\epsilon=5$ MeV
and $\bar{\epsilon}_2=19$ MeV.
 }
\label{tab:beta2_4}
\begin{tabular}{lccc}
  \hline
      $k_F$ (fm$^{-1}$) & 1.2  & 1.36 & 1.5 \\
  \hline
  $\bar{E}_0$ (MeV)  & -15.7$\div$-16.4 & -16.8  & -16.1$\div$-15.2 \\
  \hline
  $\beta^2$   & 4.5$\cdot$10$^4$  & 5.2$\cdot$10$^4$  & 7.0$\cdot$10$^4$  \\
  $\beta^2_l$ (MeV$^2$/nucleon) & 45.7  & 44.7  & 39.2  \\
  $\beta^2_u$                   & 108.6 & 91.9  & 62.7  \\
  \hline
  $S_l$       &  0.35  &  0.34  &  0.33  \\
  $S_u$       &  0.73  &  0.72  &  0.66  \\
  \hline
\end{tabular}
\end{table}

\eject

\end{document}